# Predicting the Brittle-to-Ductile Transition in Amorphous Polymers

Valeriy V. Ginzburg[1,*], Oleg Gendelman[2], and Alessio Zaccone[3]

[1]Department of Chemical Engineering and Materials Science, Michigan State University, East Lansing, Michigan, USA 48824

[2]Faculty of Mechanical Engineering, Technion, Haifa 32000003, Israel

[3]University of Milan, Department of Physics, via Celoria 16, 20133 Milano, Italy

*Corresponding author: ginzbur7@msu.edu

## Abstract

Brittle-ductile transition (BDT) is an important characteristic of amorphous (and semicrystalline) polymers. For a given strain rate, at temperatures above BDT, the polymers exhibit strain softening followed by yield and strain hardening, while at temperatures below BDT, the same materials exhibit brittle failure at relatively low strains. Surprisingly, today there is no simple model describing BDT as a function of polymer chemistry, sample history, deformation type, and strain rate. Experimental data suggest that BDT is often, though not always, associated with the β-transition. We formulate a simple scalar model to describe the visco-elasto-plastic shear stress-strain curves as functions of temperature and strain rate. We also show that within this model, there is always an upper bound on the strain rate where the material can have a uniform viscoplastic flow; this upper bound is taken to represent the BDT. We stipulate that this upper bound is inversely proportional to the Johari-Goldstein β-relaxation time. Using our "general" Sanchez-Lacombe "two-state, two-(time)scale" (SL-TS2) model, we compute the BDT for three polymers (polystyrene, poly(methylmethacrylate), and poly(vinylchloride)) and found a good agreement with experimental data.

## 1. Introduction

Amorphous polymers fail at large strains in a brittle or ductile fashion.[1] For the case of brittle failure, the stress-strain curve terminates sharply almost immediately after the elastic region. For the case of ductile failure, the stress reaches a maximum, then decreases (strain softening) and stays constant ("yield"). In many polymers, this is followed by the strain hardening due to the permanent or entanglement polymer network (as well as other factors).[2] Here, we restrict our analysis to the pre-strain-hardening region where the explicitly polymeric aspects of the problem are less pronounced. Our goal is to develop a phenomenological model describing the transition between the brittle and ductile regions in the (T, S) phase space, where T represents temperature and S corresponds to the strain rate (which could be shear, tension, or compression). This transition is known as the Brittle-Ductile Transition (BDT) and it has been the subject of extensive investigations over the years.[3–6] However, so far, theoretical understanding of the BDT is still limited. Most research revolves around the analysis of the yield stress and the brittle fracture stress as a function of T and S; the point where the two stresses are equal is identified as the BDT.[4,5] In many cases, the ductile behavior is associated with shear banding and the brittle behavior with the craze formation, though other mechanisms are also possible.[1]

From the above discussion, it follows that to estimate the BDT, one needs to calculate the stress-strain curve as a function of the strain rate and the temperature for the homogeneous system undergoing visco-elasto-plastic (VEP) deformation with yield (neglecting the strain hardening for the time being). Many models have been developed over the years, starting from Eyring[7] to Ree and Eyring[8] to Boyce and co-workers[9–11] to the "Eindhoven Glassy Polymer" (EGP) model of Govaert et al.[1,12–16] to the recent work of Rogers and co-workers,[17,18] as well as many others. While the initial models were not able to explain the "stress overshoot" (tensile or compressive stress going through a maximum before reaching the "yield plateau"), many later models were able to overcome this challenge and describe many experimental curves quite accurately and with a limited number of fitting parameters.

Recently, the fundamental premise of the Eyring approach (which is the basis of most existing VEP models) has been questioned by Long and co-workers.[19–23] According to them, the scalar description used by Eyring violates the symmetry of the problem – the "stress" used to compute the relaxation time reduction in the deformed polymers is not an invariant of a stress tensor and must be equal to zero to satisfy the symmetry requirements. They suggested instead that the reduction in the logarithm of the relaxation time must be proportional not to the stress but to the elastic energy. This analysis seemed to resolve some well-known issues with the Eyring-type approaches (e.g., the difference between the Eyring activation volume and the measured correlation volume), and new models based on it are currently being developed.

Thus, assuming that one has a good VEP model, it is possible to estimate the maximum stress that precedes the strain softening (let us denote it $\sigma_{y1}$ to differentiate from the steady-yield stress, labeled $\sigma_{y2}$; obviously, $\sigma_{y1} \geq \sigma_{y2}$). The brittle-ductile transition would take place when the stress required to initiate the brittle fracture, $\sigma_b$, is smaller than $\sigma_{y1}$. However, estimating $\sigma_b$ is even less straightforward than estimating $\sigma_{y1}$, given that fracture is inherently non-uniform and there are many modes of brittle failure. Wang and co-workers[24] provide a short review how the linear and nonlinear fracture mechanics can be used to estimate $\sigma_b$. This is a very important and broad topic, but it is outside the scope of the current paper.

In some ways, the above approach to the estimate of the BDT is similar to the search for binodals (coexistence lines) in thermodynamics.[25] The competition between the "uniform" (ductile) and "non-uniform" (brittle) failure modes is judged based on the stress or energy comparison. (Although real-life yield usually is an inhomogeneous process – see the discussion on shear banding below – it still preserves the continuity of the material and the stress uniformity throughout it). If the brittle failure mode is advantageous, the crazes or cracks will need to nucleate (or get introduced) and then grow, similar to the nucleation and growth (N&G) mechanism of crystallization or phase separation. Notably, in the same phase-separating systems, one can also find the so-called spinodal lines.[25] Spinodals represent the points in the phase space where the uniform (homogeneous) solution loses stability with respect to infinitesimally small fluctuations and becomes inherently unstable. In general, the calculation of spinodals is easier than that of binodals and is often used as a simple estimate for the phase boundary. Here, we propose to do something similar and look at the BDT as the point where a uniform VEP yield is no longer possible.

So, why are there cases where uniform post-yield flow is unsustainable? We know that for any given temperature, the transition from the ductile behavior to the brittle one occurs as the strain rate is increased. We can divide the strain into two components, the elastic strain and the viscoplastic strain. In the post-yield state, the elastic strain stays constant, while the viscoplastic strain increases without causing any change in the stress. This means that there is equilibrium between the rate of energy absorption and the rate of energy dissipation. Hence, knowing the maximum energy dissipation means knowing the BDT. As the strain rate is increased further, the material can no longer dissipate all the absorbed energy, leading to brittle failure.

For amorphous polymers, thus, the question becomes – how to calculate the maximum dissipation rate? The dissipation rate is proportional to shear viscosity, $\eta$, which, in turn, is proportional to the product of the shear modulus, $G$, and the relaxation time, $\tau$.[26] In the quiescent state, $\tau = \tau_{\alpha}\left(T\right)$, where $\tau_{\alpha}\left(T\right)$ is the $\alpha$-relaxation time usually associated with the segmental mobility of polymers.[27–31] As the material is deformed, the relaxation time decreases as the effective potential barriers between adjacent minima on the potential energy surface (PES) are reduced.[7,8,19,20,22,32–35] Yet the question remains – how much does it decrease?

It has been observed early on in the studies of the polymer BDT that there is a strong relationship between the BDT and the β-transition.[3,6,36–39] Wu in particular emphasized that the BDT can be closely related to the β-transition provided that the β-transition "is due to the main-chain motions, not due to internal motions of side groups."[39] In other words, BDT can be related to the temperature at which the β-relaxation time is on the order of the laboratory time (~$10^2$ seconds), provided that the β-relaxation is the Johari-Goldstein (JG) process.[40,41] The JG β-relaxation and the α-relaxation processes are thought to be coupled (see, e.g., Ngai and co-workers).[42–45]

Recently, we have developed the "two-state, two-(time)scale" (TS2) model describing the α- and β-relaxation times as functions of temperature for various polymeric and organic glass-formers.[46–53] Within TS2, a material is represented as a dynamic coexistence between the "solid" and "liquid" domains, with the "solid" domain fraction, $\psi$, depending on the temperature in the Boltzmann fashion. The β-relaxation time is an Arrhenius function of temperature, while the α-relaxation time merges with the β-relaxation time at high temperatures and becomes strongly super-Arrhenius at lower temperatures. Furthermore, we discovered a surprising universality that enables a prediction of both α- and β-relaxation times based on only two parameters, e.g., the glass transition temperature, $T_g$, and the fragility, $m$.[53]

Here, we utilize the new general TS2 model to estimate the brittle-ductile transition in polymers. We start by formulating a new nonlinear visco-elasto-plastic (VEP) model. We then pinpoint the "dynamic phase map" in the (T,S) space, dividing it into three regions: Type I (brittle), Type II (ductile with stress overshoot), and Type III (liquid-like behavior with no overshoot). The Type I – Type II transition is associated with BDT and expressed in terms of the α- and β-relaxation times. We thus recover the formula of Wu[6] for the BDT activation energy. We also demonstrate that our model can reasonably describe Wu's experimental data for the BDT temperature as a function of the strain rate for several amorphous polymers (poly(methylmethacrylate [PMMA], polystyrene [PS], and poly(vinylchloride) [PVC]). We conclude by discussing the limitations of the current model and propose next steps in its improvement.

## 2. The Model

### *2.1. Note on Scalar and Tensor Variables*

In this initial study, we formulate a “toy model”, sometimes labeled 1D-model (as opposed to a fully tensorial 3D model). Let us consider the scenarios where the volume is unchanged during the deformation, i.e., the trace of the deformation tensor is zero. For the shear deformation, it is satisfied automatically. For the case of pure shear,

$$\boldsymbol{\varepsilon} = \begin{bmatrix} 0 & \frac{\gamma}{2} & 0 \\ \frac{\gamma}{2} & 0 & 0 \\ 0 & 0 & 0 \end{bmatrix} \tag{1}$$

let us consider an invariant $\gamma \equiv \sqrt{2\boldsymbol{\varepsilon} : \boldsymbol{\varepsilon}}$, where $\boldsymbol{\varepsilon} : \boldsymbol{\varepsilon}$ is given by,

$$\boldsymbol{\varepsilon} : \boldsymbol{\varepsilon} \equiv \sum_{i,j=1}^{3} \left( \varepsilon_{ij} - \frac{1}{3} \varepsilon_{kk} \delta_{ij} \right)^2 \tag{2}$$

Thus, for pure shear, $\gamma = \gamma$.

For the tensile or compressive deformation, the strain tensor is given by,

$$\boldsymbol{\varepsilon} = \begin{bmatrix} -\nu\varepsilon & 0 & 0 \\ 0 & -\nu\varepsilon & 0 \\ 0 & 0 & \varepsilon \end{bmatrix} \tag{3}$$

If the Poisson’s ratio ν = 0.5, the deformation is volume-preserving, and the mechanical energy changes are, again, due to the effects of shear. The shear invariant $\gamma = \varepsilon\sqrt{3}$.

We now drop the tilde symbol and label $\gamma \equiv \sqrt{2\boldsymbol{\varepsilon} : \boldsymbol{\varepsilon}}$. The following sections are used to formulate a simple nonlinear model for the constant-strain-rate stress-strain curve expressed in terms of the “shear invariant” γ.

### *2.2. The Nonlinear Visco-Elasto-Plastic Model*

Below, we describe several main assumptions used in our analysis.

#### *2.2.1. Assumption 1 – Maxwell Element*

The first assumption is that our material can be described by a "generalized" Maxwell model. In a conventional "Maxwell element"[54,55] model, an elastic spring is connected in series with a viscous dashpot. Thus, we have,

$$\sigma_1 = \sigma_{el} = G\gamma_{el} \tag{4a}$$

$$\sigma_2 = \sigma_v = G\tau \frac{d\gamma_v}{dt} \tag{4b}$$

$$\sigma_1 = \sigma_2 = \sigma \tag{4c}$$

$$\gamma_{el} + \gamma_v = \gamma \tag{4d}$$

The use of a Maxwell element implies that the material deforms as an elastic solid at low times and then, post yield, flows as a viscous liquid. Once yielding is initiated, the stress is governed by a balance between elastic energy storage and viscous dissipation. This approximation is appropriate for describing both the early elastic regime and the post-yield regime of glassy polymers, where relaxation processes become completely activated and the material exhibits viscoplastic flow.

In case of stress relaxation experiment, in the right-hand-side of eq 4d, $\gamma = \gamma_0 = const.$, while in case of stress-strain test, $\gamma = \dot{\gamma} t$.

It is important to note that Maxwell element model implies that the elastic and the viscous stress are always equal. We will explore this property later on.

#### *2.2.2. Assumption 2 – Nonlinear Elasticity*

The elastic element described in equation 4a is a linear elastic one. As we consider larger strains, we need to stipulate that elasticity can be nonlinear. Elastic stress comes from the presence of a potential energy landscape; assuming we start from a local minimum, we can calculate the stress as a derivative of the energy with respect to strain (forgetting for now about the tensorial character of both strain and stress). Suppose that $\gamma = 0$ is a local energy minimum, and $\gamma = 2\gamma^*$ corresponds to the location of an energy barrier or maximum. Then, the stress will be equal to zero in both of those points. Let us stipulate that the nonlinear elastic stress as a function of strain is given by,[56–58]

$$\sigma = \sigma^* \sin\left(\frac{\pi}{2}\frac{\gamma_{el}}{\gamma^*}\right) = G\frac{2\gamma^*}{\pi}\sin\left(\frac{\pi}{2}\frac{\gamma_{el}}{\gamma^*}\right), \tag{5a}$$

while the strain energy density is given by,

$$U = G\left(\frac{2\gamma^*}{\pi}\right)^2\left[1-\cos\left(\frac{\pi}{2}\frac{\gamma_{el}}{\gamma^*}\right)\right] \tag{5b}$$

The maximum of elastic stress as a function of elastic strain corresponds to $\gamma_{el} = \gamma^*$, halfway between the minimum and the barrier.

#### *2.2.3. Assumption 3 – Viscous Dissipation and Relaxation Time*

Let us recall that stress relaxation of a Maxwell element is described by,

$$\tau\frac{d\gamma_{el}}{dt} + \gamma_{el} = 0 \tag{6}$$

With the initial condition, $\gamma_{el}(0) = \gamma_0$. The solution of this equation is the Debye relaxation function,

$$\frac{\gamma}{\gamma_0} \equiv Y = \exp\left[-\frac{t}{\tau}\right] \tag{7}$$

For many materials, however, relaxations are described not by the Debye function, but, e.g., by the Kohlrausch-Williams-Watts (KWW)[59–62] function,

$$Y = \exp\left[-\left(\frac{t}{\tau}\right)^{\beta}\right] \tag{8}$$

(Note that KWW function can describe relaxation only in some intermediate time range. The relaxation behavior must be Debye-like in the asymptotic limit of very small and very large times. We will discuss the reason for this requirement later).

The relaxation function is schematically shown in Figure 1.

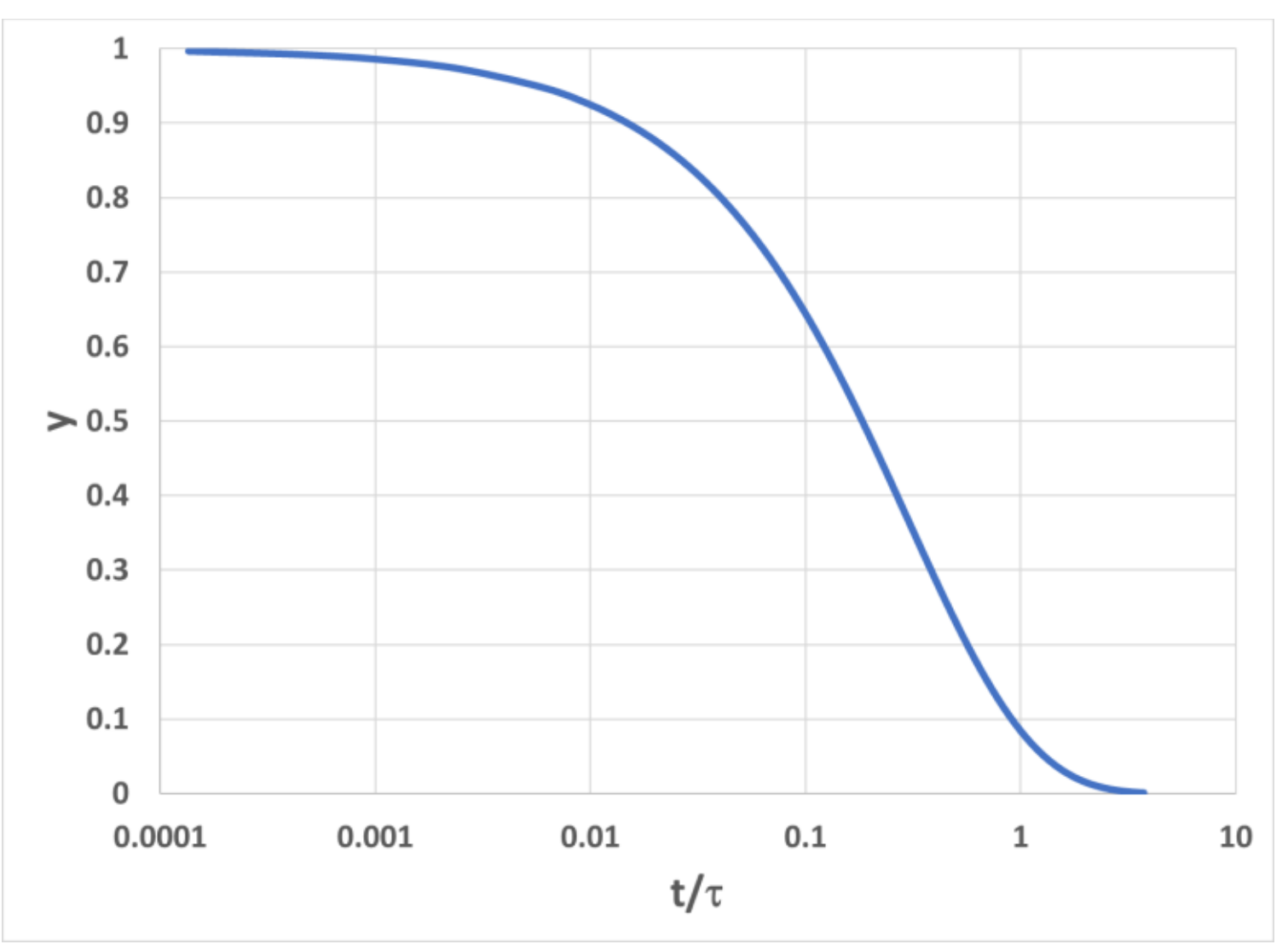

*Figure 1. Typical stress relaxation profile*

In this analysis, we assume that the KWW exponent $\beta$ = 1. Considering the cases where $\beta$ < 1 will result in an additional multiplier for the relaxation time at yield but will not fundamentally alter the results.

It has been known for many years that as the material is deformed, its relaxation time decreases. Eyring[7] justified it by proposing an analogy with activated chemical reactions – the stress is supposed to deform the potential energy landscape. The resulting formula for the dependence of the relaxation time on the stress was given by,[35]

$$\tau = \tau_0 \left[ \frac{\left( \frac{V^* \sigma}{k_B T} \right)}{\sinh\left( \frac{V^* \sigma}{k_B T} \right)} \right] \tag{9}$$

Here, $V^*$ is the "activation volume", $\sigma$ is the stress, $k_B$ is the Boltzmann's constant, and $T$ is absolute temperature.

Long and co-workers criticized the Eyring approach as flawed from the symmetry standpoint (stress here is not a true scalar and thus eq 9 is not rotationally invariant).[19,22] Instead, they proposed an alternative expression,

$$\tau = \tau_0 \exp\left[ -\frac{V^* U}{k_B T} \right] \tag{10}$$

Here, $V^*$ is the correlation volume (also known as cooperatively rearranging region [CRR]), and $U$ is the elastic strain energy density. The idea was that as the deformed material climbs up inside its energy well, the transition barrier is reduced exactly by the amount its energy is increased. One can derive equation 10 in a different, more fundamental way, by solving the governing Fokker-Planck equation, or, for diffusive dynamics, the governing Smoluchowski diffusion equation with shear strain (see, e.g., Zaccone et al.[63,64]), which generalizes the Kramers escape rate theory to systems under shear. This leads to an Arrhenius-like rate with shear-dependent activation energy, derived from the interplay of diffusion, conservative interactions, and macroscopic flow. Importantly, in this framework[63,64] the "activation volume" is not an ad hoc parameter but emerges from microscopic dynamics and from the spatial extent of the drift term and hydrodynamic coupling.

Long and co-workers assumed the quadratic ("harmonic") form for the elastic energy strain. Here we assume the more general nonlinear form (equation 5b).

### 2.2.4. Assumption 4 – Yield as the Steady State

We now simplify the problem to consider the question of interest – what determines the qualitative behavior of the stress-strain curves. The three main types of stress-strain responses (assuming a simple test in which the strain rate is kept constant) are depicted in Figure 2.

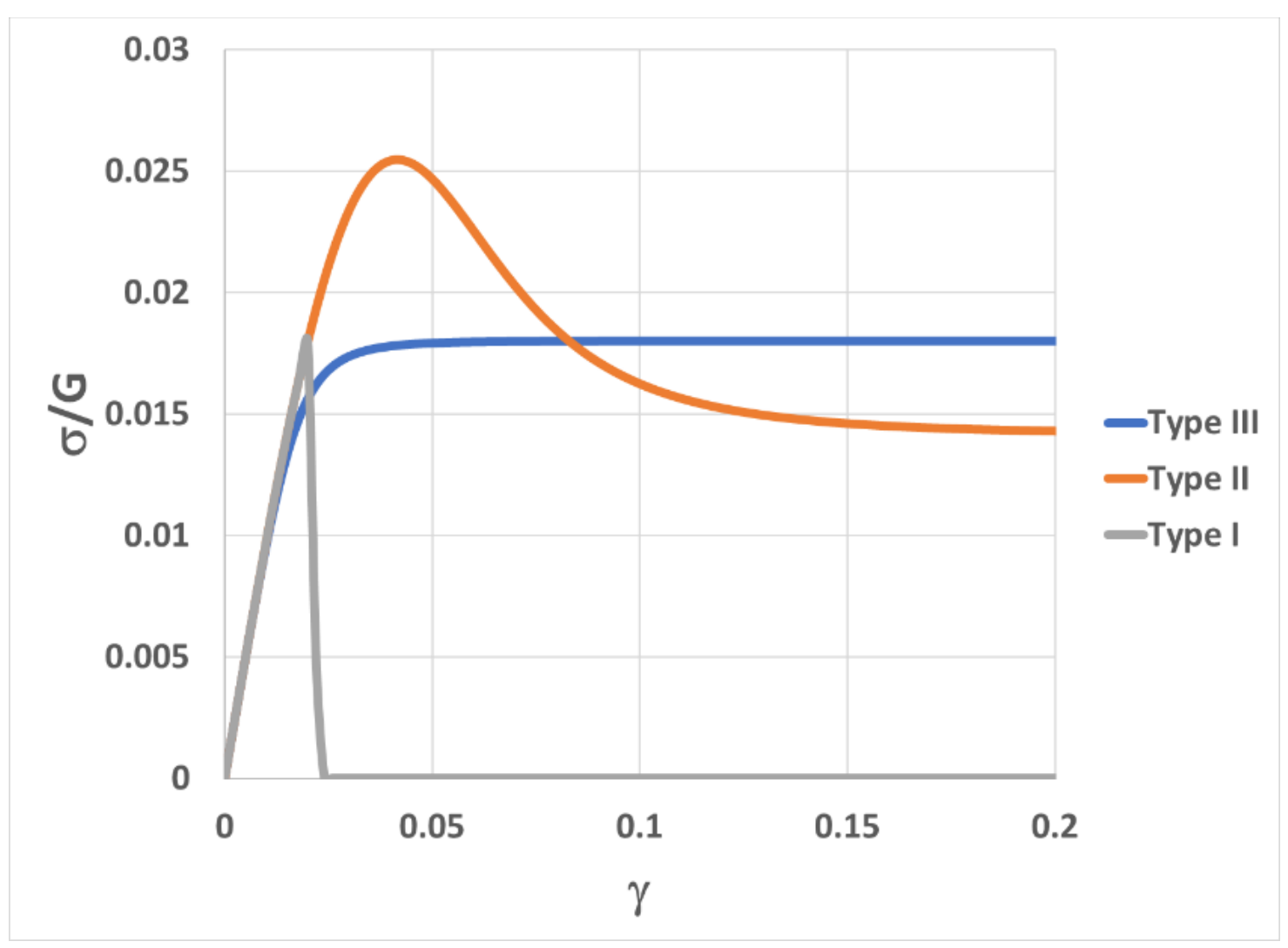

*Figure 2. Three types of stress-strain curves. Type I – brittle response; Type II – ductile response with stress overshoot; Type III – ductile response with no stress overshoot.*

The type of response (Type I – brittle; Type II – ductile with stress overshoot; Type III – ductile with no stress overshoot) is determined by the material chemistry, the thermomechanical history of the specimen, and the test temperature and shear rate. Generally, Type III corresponds to liquid or near-liquid state of material, e.g., polymer melts; Type II corresponds to soft-solid materials, e.g., polymers near their glass transition temperature; and Type I corresponds to conventional hard solids such as glass or hard thermoplastics (PS, PMMA) at room temperature. Many authors (e.g., Wang et al.,[4,5] Rogers et al.[17,18]) attempted to propose criteria for the brittle-ductile transition, but the problem is still far from being fully solved.

Let us assume that the "final yield" state is time independent. This implies that,

$$\gamma_{el} = \gamma_Y \quad (11a)$$

$$\gamma_{VP} = \gamma - \gamma_Y \quad (11b)$$

Here, $\gamma_Y$ is a constant; $\gamma_{VP}$ is the same as $\gamma_v$ in eq 1d (the label is changed to emphasize that it is "viscoplastic" shear, combining both reversible and irreversible contributions).

As discussed above, the main feature of the Maxwell model is that viscous and elastic stresses are always equal. Thus, we can consider them separately. Let us concentrate on the elastic element first. The elastic stress is given by equation 5a and plotted in Figure 3.

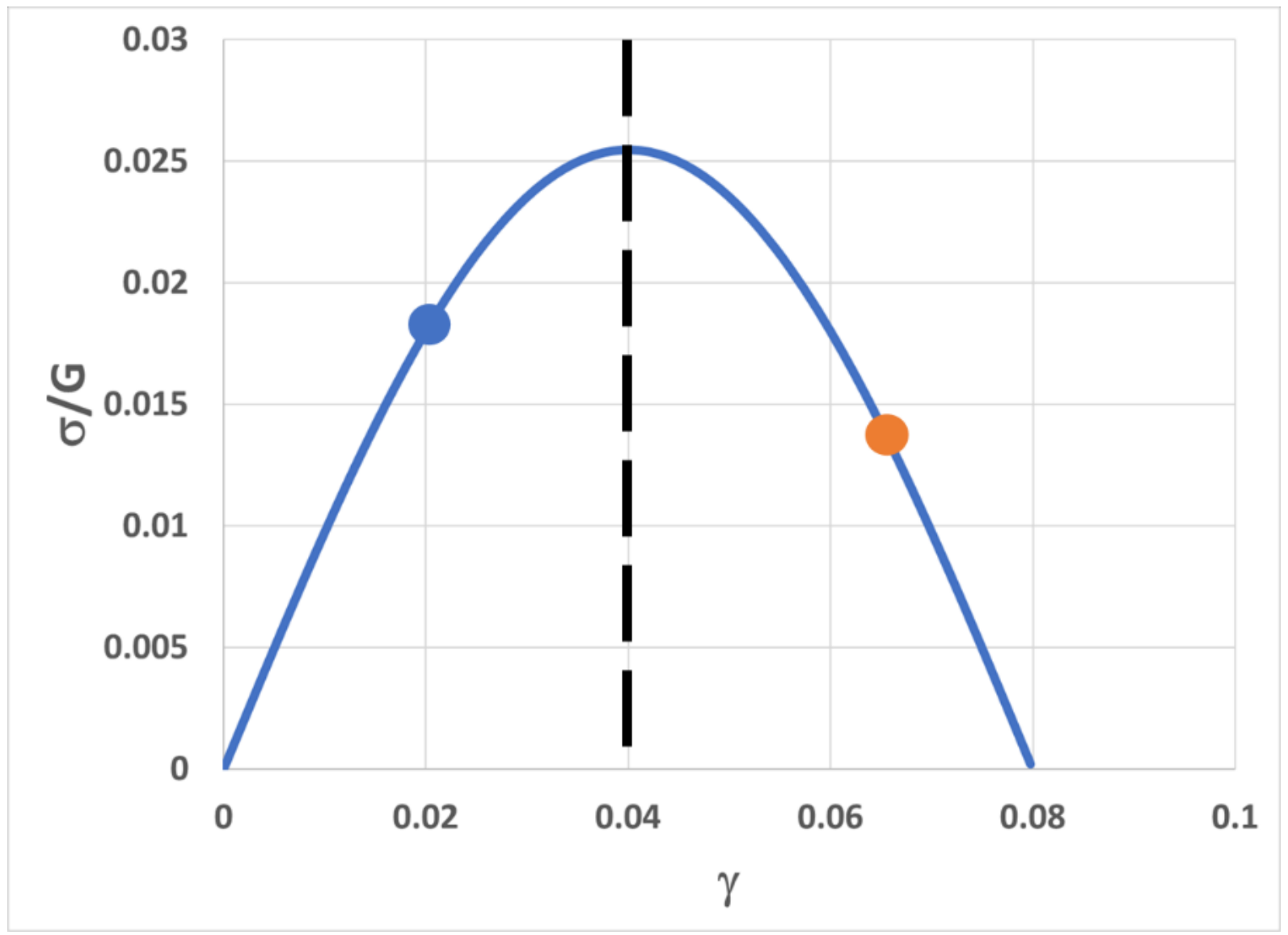

*Figure 3. Normalized elastic stress as a function of elastic strain. The vertical dashed line corresponds to the stress maximum ($\gamma^*$ = 0.04). The orange dot corresponds to the Type II curve from Figure 3 ($\gamma_Y$ = 0.067), while the blue dot corresponds to the Type III curve from Figure 3 ($\gamma_Y$ = 0.02).*

Thus, there is an obvious criterion for the transition between Type II and Type III,

$$\gamma_Y = \gamma^* \quad (12)$$

Alternatively, if we define the dimensionless variable $X = \frac{\pi}{2}\frac{\gamma_{el}}{\gamma^*}$, this criterion can be re-written as,

$$X_Y = \frac{\pi}{2} \tag{13}$$

For the time being, we postpone the question of deriving the criterion for the transition between Type II and Type I.

Let us now consider the dashpot. The viscoplastic stress is given by,

$$\sigma_{VP} = G\tau \frac{d\gamma_{VP}}{dt} \tag{14a}$$

or

$$\sigma_{VP} = G\dot{\gamma}\,\tau_\alpha(T)\exp\left[-\frac{V^* U_Y}{k_B T}\right] \tag{14b}$$

Here, the elastic energy at the yield point is,

$$U_Y = G\left(\frac{2\gamma^*}{\pi}\right)^2\left[1-\cos\left(\frac{\pi}{2}\frac{\gamma_Y}{\gamma^*}\right)\right] \tag{15}$$

The elastic stress through the yield process is, of course,

$$\sigma_{el} = G\frac{2\gamma^*}{\pi}\sin\left(\frac{\pi}{2}\frac{\gamma_Y}{\gamma^*}\right) \tag{16}$$

Equating the elastic stress (equation 16) and the viscoplastic stress (equation 14b) enables us to solve for the unknown $\gamma_Y$.

This concludes the model description. We now discuss the results. First, we present the non-dimensional (material-independent) phase diagram. Then, we apply the prediction to three amorphous polymers (PMMA, PS, and PVC) and compare the modeling results with experimental data compiled by Wu.[6]

# 3. Results and Discussion

## 3.1. *The Phase Diagram*

The equation for $\gamma_Y$ is,

$$\frac{2\gamma^*}{\pi}\sin\left(\frac{\pi}{2}\frac{\gamma_Y}{\gamma^*}\right)=\dot{\gamma}\,\tau_\alpha\left(T\right)\exp\left[-\frac{V^*}{k_BT}G\left(\frac{2\gamma^*}{\pi}\right)^2\left[1-\cos\left(\frac{\pi}{2}\frac{\gamma_Y}{\gamma^*}\right)\right]\right]$$

(17)

Denoting $X_Y=\frac{\pi}{2}\frac{\gamma_Y}{\gamma^*}$, $B=\frac{V^*G}{k_BT}\left(\frac{2\gamma^*}{\pi}\right)^2$, $Q=\frac{\pi}{2\gamma^*}$, and $Y=Q\dot{\gamma}\,\tau_0$, we finally obtain,

$$F\left(X_Y\right)\equiv\exp\left(B\left[1-\cos\left(X_Y\right)\right]\right)\sin\left(X_Y\right)=Y \qquad (18)$$

Equation 18 is the last step in the derivation. We are now looking to explore the phase diagram on the *(B,Y)* plane by investigating its solutions. Let us plot $F\left(X_Y\right)$ for several different values of *B* (Figure 4).

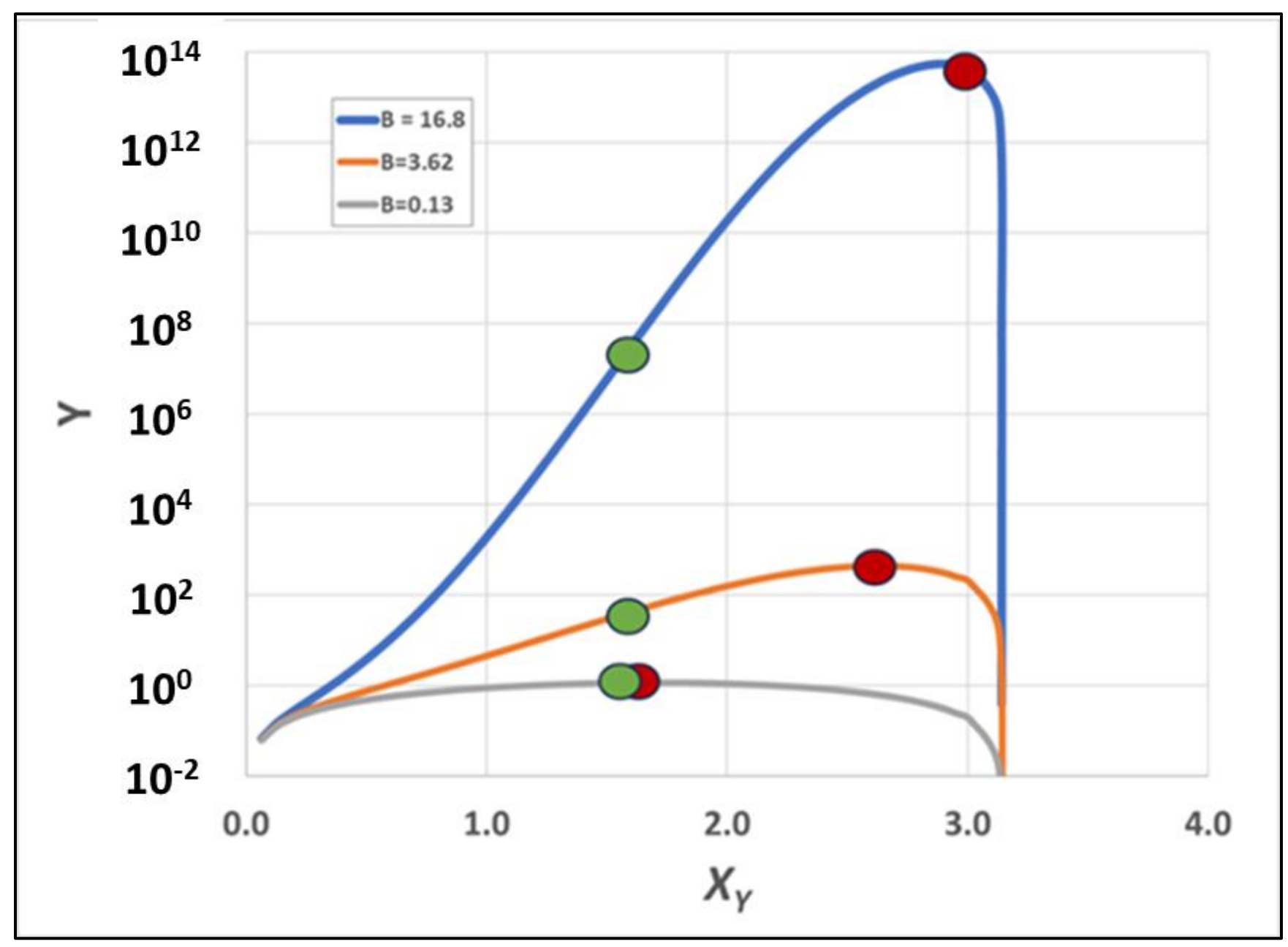

*Figure 4. The "determinant function" $F(X_Y)$ for several different values of B (see legend). The red circles correspond to the phase boundary between Type I and Type II. The green circles correspond to the boundary between Type II and Type III. For small values of B, the two merge and the Type II behavior is no longer seen.*

Depending on Y, equation 18 can have two, one, or zero solutions. Graphically, those solutions are points where the graph of the function *F* crosses a particular horizontal line. For small Y, there are two solutions – one stable (left) and one unstable (right). At some point (indicated by the red circle), the two solutions merge. Let us denote this point as $Y_1(B)$. For $Y > Y_1$, equation 18 admits of no real solutions. We can interpret this as a sign that the material is unable to plastically flow for those high shear rates. Thus, the condition $Y = Y_1(B)$ can be identified as the ductile-brittle transition.

In the region where the plastic flow is possible, $Y < Y_1$, we need to differentiate between the Type II (ductile with stress overshoot) and Type III (ductile with no stress overshoot). The transition between those two scenarios occurs, as discussed above, when $X_Y = \frac{\pi}{2}$ (eq 13). This is depicted by the green circles, denoted $Y_2(B)$. As *B* is decreased (e.g., by reducing the temperature or reducing the correlation volume or reducing the modulus), the gap between the two transition narrows and eventually disappears.

Repeating the calculations for multiple values of B, we can plot $Y_1(B)$ and $Y_2(B)$ (Figure 5). The region above the $Y_1(B)$ line is that corresponding to the brittle fracture (Type I), the region between $Y_1(B)$ and $Y_2(B)$ lines corresponds to the yield with stress overshoot (Type II), and the region below the $Y_2(B)$ line is the region with no stress overshoot (Type III).

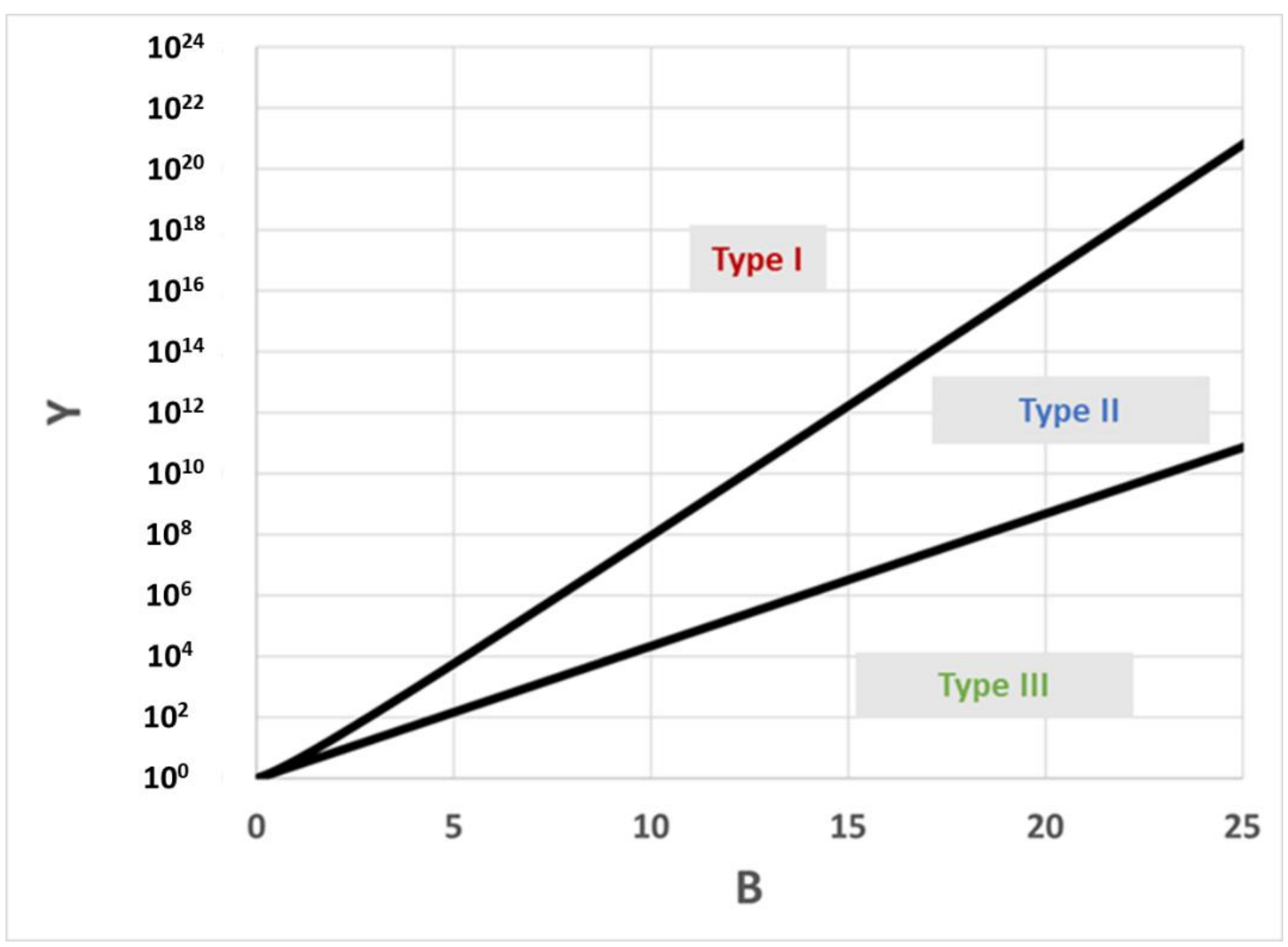

*Figure 5. Phase diagram describing different stress-strain regimes. Here, Type I refers to the brittle response; Type II corresponds to the ductile response with stress overshoot; and Type III is the yield with no overshoot.*

The Type II region is practically non-existent for small *B*. The parameter B calibrates the strength of the “acceleration due to deformation”. It depends on:

- The half-width of the potential well, $\gamma^*$;
- The correlation volume of the material, $V^*$;
- The temperature, $T$;
- The modulus, $G$.

In the future, we can investigate the specific materials under various test conditions and place them on this phase diagram.

### *3.2. Calculating the Brittle-Ductile Transition for Polymers*

So far, we described the brittle-ductile transition in terms of B and Y. Let us now see how this analysis can be applied to real-life polymers. The brittle-ductile transition line (Figure 5) can be approximated as,

$$\ln\left[\frac{\pi}{2\gamma^*}\dot{\gamma}\,\tau_\alpha(T)\right] = kB \tag{19}$$

Here, k ≈ 1.88 is simply the result of numerical fitting to an exponential form – there is no physical significance to this number. In writing equation 19, we replaced $\tau_0$ with $\tau_\alpha(T)$ to emphasize that the α-relaxation time is the appropriate timescale for our Maxwell element. However, we still need to determine B.

Let us now make the following assumption. In equation 18, the maximum "Eyring" relaxation time reduction is *exp(2B)*. However, it is clear that the maximum relaxation time reduction in a glassy polymer must be equal to $\tau_\alpha(T)/\tau_\beta(T)$, and the lower bound for the relaxation time is $\tau_\beta(T)$, where $\tau_\beta(T)$ is the Johari-Goldstein β-relaxation time. This is because even if all possible vestiges of the glassy state were eliminated and the material became a deeply supercooled liquid, its characteristic time would be $\tau_\beta(T)$. In reality, of course, the reduction can be smaller, and the relaxation time would be larger. Let us write,

$$2B = x\ln\left(\frac{\tau_\alpha(T)}{\tau_\beta(T)}\right) \tag{20}$$

Here $0 \le x \le 1$ is a new dimensionless parameter describing the efficiency of the softening process. For example, in materials with smaller correlation volume, the dissipation becomes less effective than in those with larger correlation volume. Thus, we can say that when $x \approx 0$, there can be almost no strain softening and the material is either liquid (Type III, slow shear) or brittle (Type I, fast shear). On the other hand, when $x \approx 1$, we expect to see the Type II behavior (stress overshoot followed by yield) at temperatures below $T_g$ in a broad range of shear rates. Thus, in some ways, there is a similarity between the parameter *x* and the "brittility" factor introduced by Rogers and co-workers,[17,18] although it is not a one-to-one correspondence.

Substituting eq 20 into eq 19, we obtain,

$$\ln\left[\frac{\pi}{2\gamma^*}\dot{\gamma}\,\tau_\alpha(T)\right]=\frac{kx}{2}\ln\left(\frac{\tau_\alpha(T)}{\tau_\beta(T)}\right) \quad (21a)$$

or

$$\ln\left[\frac{\pi}{2\gamma^*}\dot{\gamma}\right]=-\Phi\ln\left(\tau_\beta(T)\right)-\left(1-\Phi\right)\ln\left(\tau_\alpha(T)\right) \quad (21b)$$

where $\Phi=\frac{kx}{2}$.

If one assumes that the temperature range of interest is narrow and can be approximated by the Arrhenius function even for the α-relaxation time, we recover the phenomenological equation of Wu[6]

$$E_b=\Phi E_\beta+\left(1-\Phi\right)E_\alpha \quad (22)$$

Here, $E_b$ is the activation energy for the brittle-ductile transition (governing the dependence of the brittle-ductile strain rate on temperature or vice versa), $E_\beta$ is the activation energy of the JG process (and, according to the "classical" scenario,[65] it is also the activation energy for the liquid flow at high temperatures), while $E_\alpha$ is the apparent activation energy of the $\alpha$-process at temperatures slightly below $T_g$. Equation 22 can be interpreted as a "mixing rule", describing the relative contributions of the α- and β-processes in setting the sensitivity of the BDT to temperature or shear rate. This will be discussed below when we compare the behavior of different polymers.

Within TS2,

$$\ln\left(\tau_\beta(T)\right)=\ln\left(\tau_{el}\right)+\overline{E_1}\left(\frac{T_X}{T_g}\right)\left(\frac{T_g}{T}\right) \quad (23a)$$

$$\ln\left(\tau_\alpha\left(T\right)\right)=\ln\left(\tau_{el}\right)+\overline{E_1}\left(\frac{T_X}{T_g}\right)\left(\frac{T_g}{T}\right)+\left(\overline{E_2}-\overline{E_1}\right)\left(\frac{T_X}{T_g}\right)\left(\frac{T_g}{T}\right)\psi\left(T\right)$$

(23b)

Here, $T_X$ is the thermodynamic glass-liquid transition, and $\overline{E_1}=E_1/\left(k_B T_X\right)\approx 50$ and $\overline{E_2}=E_2/\left(k_B T_X\right)\approx 150$ are the universal non-dimensional activation energies of the liquid and solid states, respectively.[53] The ratio $\left(\frac{T_X}{T_g}\right)$ and the elementary time, $\tau_{el}$, depend on the polymer fragility, *m*. The solid fraction, $\psi$, is given by,

$$\psi\left(T\right)=\begin{cases}\dfrac{1}{1+\exp\left(\Delta S\left[1-\left(\dfrac{T_X}{T_g}\right)\dfrac{T_g}{T}\right]\right)}, T\geq T_g\\ \psi_g+p\left(1-\dfrac{T}{T_g}\right), T<T_g\end{cases} \qquad (24)$$

Here, the parameter $\Delta S=16.0$ is universal, while $\left(\frac{T_X}{T_g}\right)$ and $\psi_g$ are the functions of the fragility. The solid fraction at low temperatures is non-equilibrium and can be approximated as a linear function of temperature. The parameter *p* can be estimated on the basis of the general SL-TS2 theory[53] combined with the Simha-Boyer[66] and Boyer-Spencer[67,68] rules for the thermal expansion of polymers. Within SL-TS2, we estimated the relative volume difference between the liquid and the solid states of the material to be approximately $2\xi\cong 0.07$. The "Boyer rules" suggest $\alpha_g T_g\cong 0.06\pm 0.02$. Within SL-TS2, below the glass transition, all the change in density occurs due to the change in the solid fraction, $\psi$. Thus, $p=\frac{\alpha_g T_g}{2\xi}$. In our calculations below, we set *p* = 0.9.

The above analysis was done for the "fresh" polymers; in reality, polymer properties change due to aging. If a glassy polymer has been aged at temperature $T < T_g$ for some time $t_a$, we expect that its $\alpha$-relaxation time has increased and become comparable to $t_a$.[27] In that case, we can evaluate the "fictive temperature" and then re-calculate $\psi(T)$ as,

$$\psi(T) = \psi_f + p\left(\frac{T_f}{T_g} - \frac{T}{T_g}\right) \tag{25}$$

where $T_f$ and $\psi_f$ are the solutions of the following set of equations,

$$\psi_f = \frac{1}{1+\exp\left(\Delta S\left[1-\left(\frac{T_X}{T_g}\right)\frac{T_g}{T_f}\right]\right)} \tag{26a}$$

$$\ln(t_a) = \ln(\tau_{el}) + \overline{E_1}\left(\frac{T_X}{T_g}\right)\left(\frac{T_g}{T_f}\right) + \left(\overline{E_2} - \overline{E_1}\right)\left(\frac{T_X}{T_g}\right)\left(\frac{T_g}{T_f}\right)\psi_f \tag{26b}$$

In our analysis below, we assume that the polymers were aged for about 1 month, thus $\log(t_a) = 6.42$. The fictive temperatures and fictive solid fractions will be calculated and tabulated for each of the three polymers.

### *3.3. Comparison with Experiment – BDT for PS, PMMA, and PVC*

It is important to note, before attempting to compare our model with the data, that the experimental stress-strain curves depend significantly on multiple factors. Those factors include: (1) polymer molecular weight and polydispersity; (2) impurities; (3) sample history (including aging at temperatures below $T_g$); (4) precision of the measurement instruments. Thus, to illustrate just one example, Hu et al.[69] show PMMA exhibiting Type II ductile behavior under compression at room temperature (T = 25 °C), while Baer and co-workers[3] used tensile experiments to observe the brittle-ductile transition at T = 42 °C.

The analysis of Wu[6] is based on tensile experiments at various strain rates. As discussed above, we need to relate the effective shear strain and strain rate to the experimentally observed tensile strain and strain rate – it is $\dot{\gamma} \approx \dot{\varepsilon}\sqrt{3}$, and $\gamma^* \approx \varepsilon^* \sqrt{3}$. We will use $\varepsilon^* \approx 0.06$, which is reasonably close to the yield or fracture strain reported in the study.

The brittle-ductile transition temperatures are calculated by varying $\dot{\varepsilon}$ between 1.69x10$^{-5}$ and 1.69x10$^{-2}$ s$^{-1}$ (corresponding to the range studied by Wu[6]) and numerically solving equations 21b, 23a-23b, 25, and 26a-26b. The only adjustable parameter here is *x* – all the other parameters are taken from our earlier TS2 papers where the material parameters for PS, PMMA, and PVC have been determined based on the dielectric and volumetric data. The model results and the experimental data are plotted in Figure 6 for PMMA, Figure 7 for PS, and Figure 8 for PVC. The lines are the model predictions, the symbols are data from Wu,[6] and the error bars are 5 K.

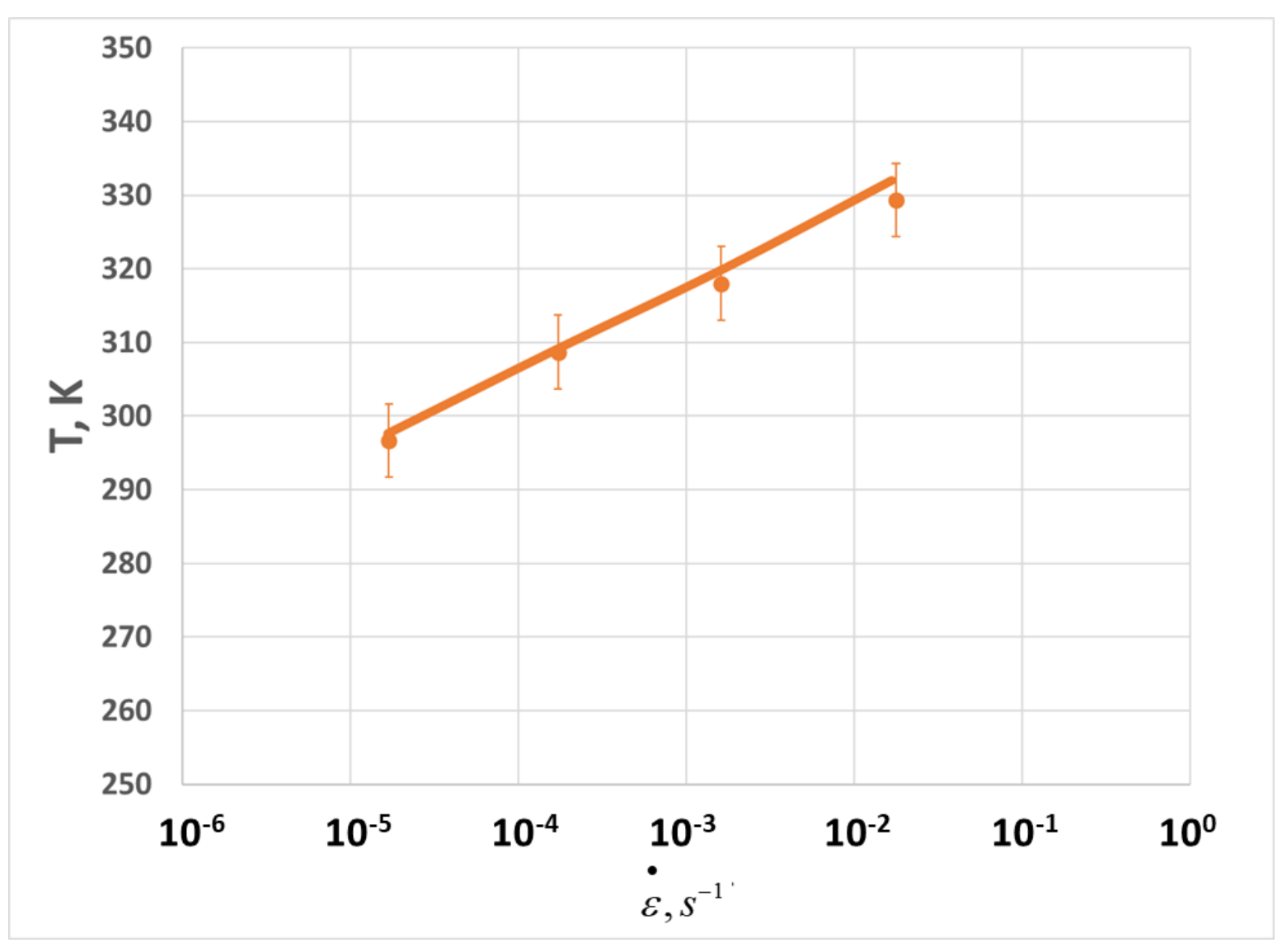

*Figure 6. Brittle-ductile transition for PMMA. Circles are the data from Wu,[6] line is the SL-TS2 model estimate.*

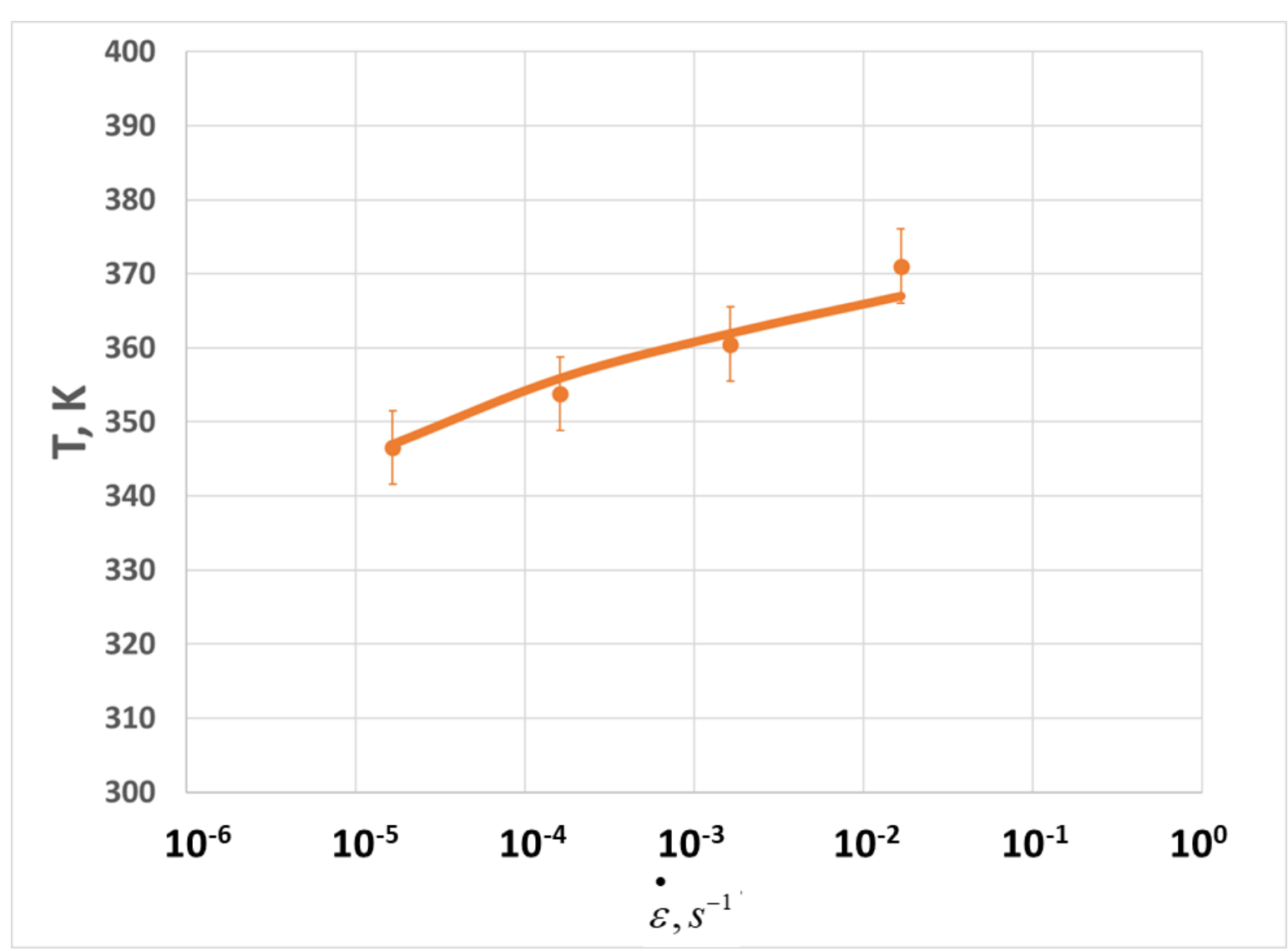

*Figure 7. Brittle-ductile transition for PS. Circles are the data from Wu,[6] line is the SL-TS2 model estimate.*

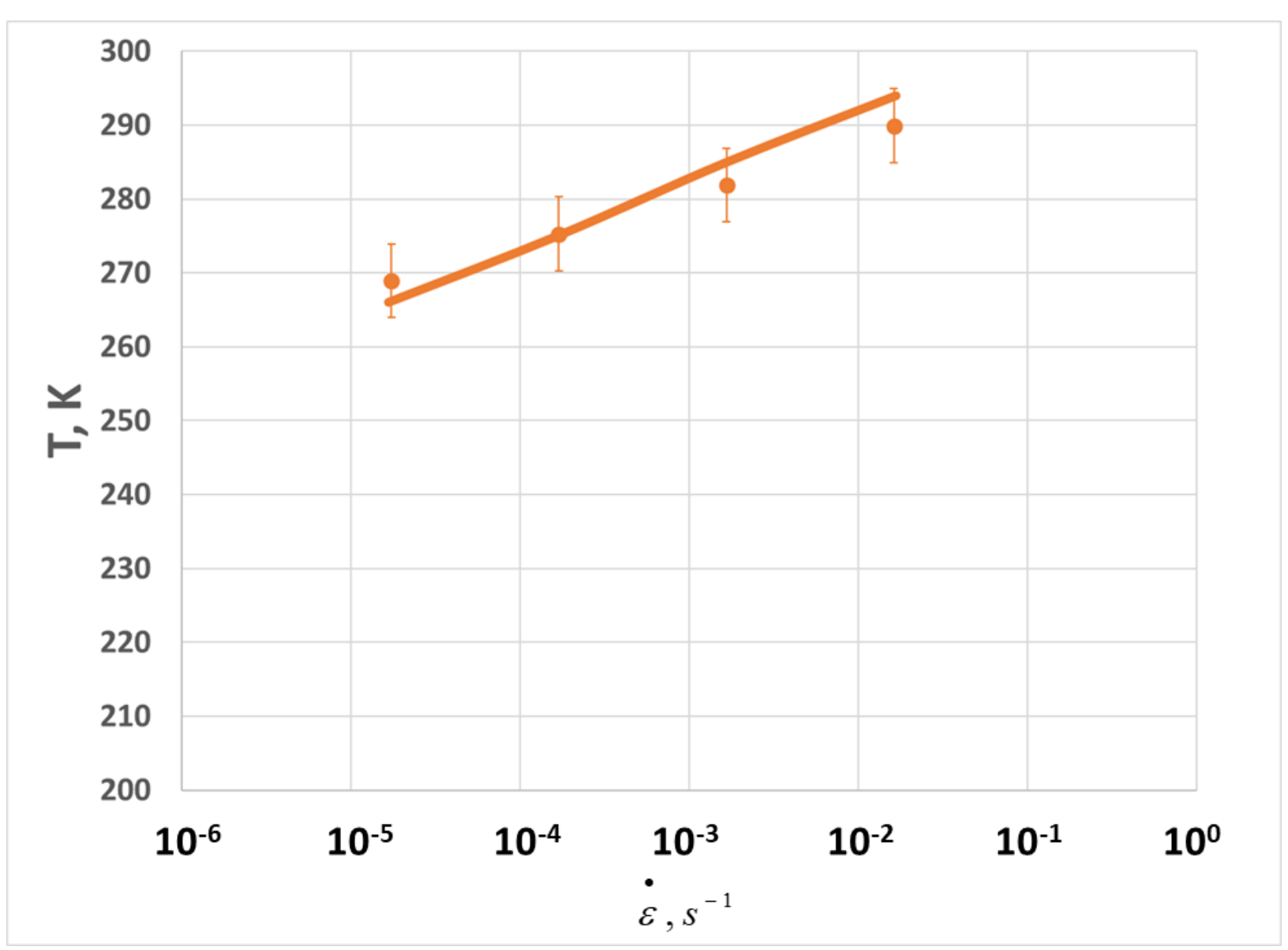

*Figure 8. Brittle-ductile transition for PVC. Circles are the data from Wu,[6] line is the SL-TS2 model estimate.*

The model parameters are shown in Tables 1 and 2. Table 1 contains the TS2 parameters – they were regressed in our earlier study[53] and are used without any modification here. Table 2 has the fictive temperature and the corresponding value of $\psi$ (estimated using equations 25 and 26a-26b), the fitting parameter $x$, the calculated brittle-ductile transition temperature at the strain rate $\dot{\varepsilon} = 1.67 \bullet 10^{-2} s$, the calculated brittle-ductile activation energy (in kcal/mol) and the experimentally measured brittle-ductile activation energy.[6] The agreement is quite reasonable given the simplicity of the model.

*Table 1. TS2 model parameters for PMMA, PS, and PVC. See text for more details.*

| Polymer | $T_g$, K | m | $T_x$, K | $\psi_g$ | $\log(\tau_{el})$ |
|---|---|---|---|---|---|
| PMMA | 378 | 100 | 339 | 0.159 | -23.6 |
| PS | 373 | 100 | 334 | 0.159 | -23.6 |
| PVC | 350 | 120 | 320 | 0.201 | -25.8 |

*Table 2. Parameters describing the brittle-ductile transition. See text for more details.*

| Polymer | $T_f$, K | $\psi_f$ | $x$ | $T_{b1}$, K | $E_b$, kcal/mol | $E_{b,exp}$, kcal/mol |
|---|---|---|---|---|---|---|
| PMMA | 363.5 | 0.252 | 0.95 | 332 (±3) | 40 (±5) | 44 |
| PS | 358.7 | 0.252 | 0.49 | 367 (±3) | 85 (±5) | 77 |
| PVC | 336.5 | 0.312 | 0.94 | 294 (±3) | 40 (±5) | 54 |

The most interesting result is the difference in behavior between PS, on the one hand, and PMMA and PVC, on the other. The analysis of Wu[6] shows that the BDT in PS is very close to $T_g$, while in PMMA and PVC it is closer to the β-transition. In our model, it is represented by the "efficiency" parameter *x*. It can be seen that *x* is close to 1 for PMMA and PVC, but is only about 0.5 for PS. In other words, when the system is close to the top of the potential energy barrier, the solid fraction is nearly zero for PMMA and PVC, and the material is essentially a deeply supercooled liquid. For PS, on the other hand, even at this point, there is a significant fraction of solid domains.

## 4. Discussion

The behavior of the brittle-ductile transition as a function of polymer structure, sample history, temperature, and strain rate is complex and non-trivial. The classical approach to predicting BDT is to estimate the yield stress (corresponding to the ductile failure) and the brittle fracture stress (corresponding to the brittle failure). The measurement or evaluation of the latter is typically done within the framework of fracture mechanics.[4,24,70] The criterion for BDT is that the brittle failure stress equals the yield stress; at lower temperatures or higher strain rates, the preferred failure mechanism is brittle, while at higher temperatures or lower strain rates, it is ductile. In some ways, this is similar to the determination of a phase coexistence line in thermodynamics.

Our assumption here is that there is a different way to estimate the BDT which hitherto has not been widely used. The yield flow can be thought of as a uniform process (neglecting the difference between the "neck" and the rest of the sample in a tensile experiment). Suppose that such a uniform process loses its stability as the strain rate is increased (at constant temperature). Thus, if we can find the criterion for the loss of stability of the uniform yield flow, we will automatically obtain an estimate for BDT – no need for explicit fracture mechanics calculations. This would be similar to finding a spinodal instead of a phase coexistence (binodal). The calculation of a spinodal usually does not provide an exact phase coexistence criterion, but it is often good enough and does not require a detailed free energy calculation of a non-uniform phase. We are proposing the same approach to the BDT problem.

Predicting the stability of a viscoelastoplastic (VEP) stress-strain curve thus requires a reasonably accurate VEP model. There are many VEP models available[1,9–12,14,15,17,18,20,21,23] (this is not an exclusive list by any means!), yet only some of them allow for the loss of stability at high strain rates. By replacing the harmonic elastic strain energy density with the Frenkel-style cosine function,[56,57] we ensure that the relaxation time decrease has an upper bound. This, in combination with the use of the Maxwell element model, leads to the "phase diagram" of Figure 5, distinguishing three possible regions: Type I (brittle failure), Type II (stress-strain curve with stress overshoot, strain softening, and yield), and Type III (stress-strain curve with yield and no overshoot). The phase space has two independent variables –

Y (normalized strain rate) and B (normalized inverse temperature). For small values of B (high temperatures or small correlation volume), the Type II behavior is squeezed out and there is a direct transition from liquid-like flow to solid-like brittle failure.

To compare our predictions with experiments, it is necessary to relate the scaled variables Y and B to experimentally measured strain rate and temperature. To do this, we need to know the relaxation times (both $\alpha$- and $\beta$-) and the correlation volume. The relaxation times are computed based on the SL-TS2 theory; we note that the $\alpha$-relaxation time depends on the aging (and possibly other conditions). The correlation volume is more difficult to estimate. Instead of calculating the correlation volume based on first principles, we introduce the "efficiency" parameter *x*, describing the maximum fraction of the solid elements that "melt" as the system approaches the steady-state yield. For PMMA and PVC, $x \sim 1$, indicating that the relaxation time drops very significantly as the material is deformed. For PS, on the other hand, $x \sim 0.5$, and the relaxation time (or viscosity) of the flowing polymer is significantly larger. This results in PS being more brittle relative to PMMA, in agreement with the data. Other polymers, like polycarbonate, exhibit significantly less brittle behavior and have very low brittle-ductile temperatures. Describing those polymers will likely require a more in-depth analysis of the $\beta$-relaxation process itself, which will be one of our future tasks.

A useful perspective on the loss of stability of homogeneous viscoplastic flow proposed here can be drawn from the extensive literature on shear banding in soft glassy materials.[71–74] In particular, Bonn and co-workers[71] demonstrated that shear banding can arise even under homogeneous stress conditions as a consequence of an intrinsic flow instability associated with a non-monotonic steady-state flow curve. In their framework, the competition between structural buildup (aging or aggregation or formation of new links) and flow-induced breakdown leads to a region where the derivative of stress with respect to strain rate becomes negative, resulting in coexistence between flowing and arrested regions.

The present model can be viewed as a limiting case of the same general mechanism. Here, the decrease of the relaxation time with deformation is bounded from below by the Johari–Goldstein β-relaxation time, which imposes a maximum rate of energy dissipation. As the strain rate increases, this bound prevents further softening, leading to the disappearance of steady-state viscoplastic solutions beyond a critical rate. In contrast to shear banding, where the instability is resolved through spatial coexistence of phases, the system in the present case is unable to sustain any homogeneous flow state and instead undergoes brittle failure. In this sense, the brittle–ductile transition identified here can be interpreted as an extreme manifestation of the same class of flow instabilities that give rise to shear localization in soft glassy materials, with the key difference being the existence of a fundamental lower bound on the relaxation time.

## 5. Conclusions

We developed a simple model to predict the brittle-ductile transition in amorphous polymers. The model is based on the idea that the BDT can be visualized as the failure of the system to dissipate absorbed energy fast enough to maintain a uniform flow. We use a simple nonlinear expression for the elastic strain energy density and combined it with the TS2 estimate for the $\alpha$- and $\beta$-relaxation times. As a result, we obtained a phase map delineating the different modes of failure (brittle failure; yield with stress overshoot; yield without stress overshoot) and used it to estimate the brittle-ductile transition for three amorphous polymers (PS, PMMA, and PVC). The results agreed well with experimental data.

While the initial results are encouraging, there are a number of ways the model can be further verified, refined and improved. First, it is known that BDT is sensitive to aging. Second, while the current model assumes no volume change, in real life the bulk contributions could play a role in addition to the shear stresses. Finally, although the model here has been used only to predict BDT, it can also be applied to model the actual stress-strain curves (at least prior to strain hardening). These will be the subject of future research.

## 6. Acknowledgments

V.G. thanks Prof. Eric Baer (CWRU) and Prof. Shi-Qing Wang (University of Akron) for helpful suggestions.